\pdfoutput=1

\documentclass[12pt,a4paper]{article}

\usepackage{float}
\usepackage{ifthen} 
\usepackage[
  separate-uncertainty=true,
  table-align-uncertainty=true,
]{siunitx}
\usepackage{tabularx}
\usepackage{booktabs}
\usepackage{upgreek}

\newboolean{pdflatex}
\setboolean{pdflatex}{true} 

\newboolean{articletitles}
\setboolean{articletitles}{true} 

\newboolean{uprightparticles}
\setboolean{uprightparticles}{false} 

\def\paperauthors{LHCb collaboration} 
\def\paperasciititle{Multidifferential study of identified charged hadron distributions in Z-tagged jets in proton-proton collisions at sqrt{s}=13~TeV} 
\def\papertitle{Multidifferential study of identified charged hadron distributions in \Z-tagged jets in proton-proton collisions at $\sqrt{s}=$13~TeV} 
\def\paperkeywords{{High Energy Physics}, {LHCb}} 
\def\papercopyright{\the\year\ CERN for the benefit of the LHCb collaboration} 
\def\paperlicence{CC BY 4.0 licence}
\def\paperlicenceurl{https://creativecommons.org/licenses/by/4.0/}

\usepackage[top=1in, bottom=1.25in, left=1in, right=1in]{geometry}
\usepackage{microtype}
\usepackage{lineno}  
\usepackage{xspace} 
\usepackage{caption} 

\usepackage{graphicx}  
\usepackage{color}
\usepackage{colortbl}
\graphicspath{{./figs/}} 

\usepackage{amsmath} 
\usepackage{amssymb}
\usepackage{amsfonts}
\usepackage{upgreek} 

\newcommand*\patchAmsMathEnvironmentForLineno[1]{%
\expandafter\let\csname old#1\expandafter\endcsname\csname #1\endcsname
\expandafter\let\csname oldend#1\expandafter\endcsname\csname
end#1\endcsname
 \renewenvironment{#1}%
   {\linenomath\csname old#1\endcsname}%
   {\csname oldend#1\endcsname\endlinenomath}%
}
\newcommand*\patchBothAmsMathEnvironmentsForLineno[1]{%
  \patchAmsMathEnvironmentForLineno{#1}%
  \patchAmsMathEnvironmentForLineno{#1*}%
}
\AtBeginDocument{%
\patchBothAmsMathEnvironmentsForLineno{equation}%
\patchBothAmsMathEnvironmentsForLineno{align}%
\patchBothAmsMathEnvironmentsForLineno{flalign}%
\patchBothAmsMathEnvironmentsForLineno{alignat}%
\patchBothAmsMathEnvironmentsForLineno{gather}%
\patchBothAmsMathEnvironmentsForLineno{multline}%
\patchBothAmsMathEnvironmentsForLineno{eqnarray}%
}

\newcommand{\lhcborcid}[1]{\href{https://orcid.org/#1}{\hspace*{0.1em}\raisebox{-0.45ex}{\includegraphics[width=1em]{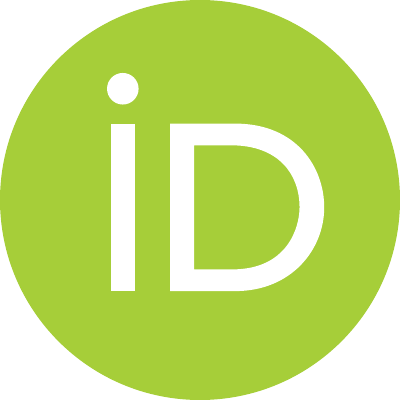}}}}

\usepackage{hyperxmp}
\usepackage[pdftex,
            pdfauthor={\paperauthors},
            pdftitle={\paperasciititle},
            pdfkeywords={\paperkeywords},
            pdfcopyright={Copyright (C) \papercopyright},
            pdflicenseurl={\paperlicenceurl}]{hyperref}
\usepackage[colorinlistoftodos,textsize=scriptsize]{todonotes}
\usepackage[bottom,flushmargin,hang,multiple]{footmisc}
\usepackage[all]{hypcap} 
\usepackage{cite} 
\usepackage{mciteplus}

\def\jt         {\ensuremath{j_{\mathrm{T}}}\xspace}
\def\z         {\ensuremath{z}\xspace} 
\def\pjet       {\ensuremath{\mathbf{p}_{\mathrm{jet}}}\xspace}
\def\phad       {\ensuremath{\mathbf{p}_{\mathrm{had}}}\xspace}
\def\Nhad       {\ensuremath{N_{\mathrm{had}}}\xspace}
\def\PScal      {\ensuremath{\mathcal{P\!S}}\xspace}
\def\zjet       {\ensuremath{\Z+\mathrm{jet}}\xspace}
\def\p         {\ensuremath{p}\xspace} 
\def\sqs   {\ensuremath{\protect\sqrt{s}}\xspace}
\def\lhcb   {\mbox{LHCb}\xspace}
\def\W      {\ensuremath{W}\xspace}     
\def\Z      {{\ensuremath{Z}}}

\def\Pp      {\ensuremath{p}\xspace}       
\def\proton      {{\ensuremath{\Pp}}\xspace}       
\def\PK      {\ensuremath{K}\xspace}    
\def\kaon    {{\ensuremath{\PK}}\xspace}
\def\Ppi         {\ensuremath{\pi}\xspace}  
\def\pion   {{\ensuremath{\Ppi}}\xspace}
\def\pipm   {{\ensuremath{\pion^\pm}}\xspace}
\def\Kpm     {{\ensuremath{\kaon^\pm}}\xspace}
\def\pt         {\ensuremath{p_{\mathrm{T}}}\xspace}
\def\pythia     {\mbox{\textsc{Pythia}}\xspace}
\def\evtgen     {\mbox{\textsc{EvtGen}}\xspace}
\def\photos     {\mbox{\textsc{Photos}}\xspace}
\def\geant      {\mbox{\textsc{Geant4}}\xspace}

\def\deriv {\ensuremath{\mathrm{d}}}
\newcommand{\aunit}[1]{\ensuremath{\text{\,#1}}}   
\newcommand{\tev}{\aunit{Te\kern -0.1em V}\xspace}
\newcommand{\eg}{\mbox{\itshape e.g.}\xspace}

\begin{document}

\renewcommand{\thefootnote}{\fnsymbol{footnote}}
\setcounter{footnote}{1}
\begin{titlepage}
\pagenumbering{roman}

\vspace*{-1.5cm}
\centerline{\large EUROPEAN ORGANIZATION FOR NUCLEAR RESEARCH (CERN)}
\vspace*{1.5cm}
\noindent
\begin{tabular*}{\linewidth}{lc@{\extracolsep{\fill}}r@{\extracolsep{0pt}}}
\ifthenelse{\boolean{pdflatex}}
{\vspace*{-1.5cm}\mbox{\!\!\!\includegraphics[width=.14\textwidth]{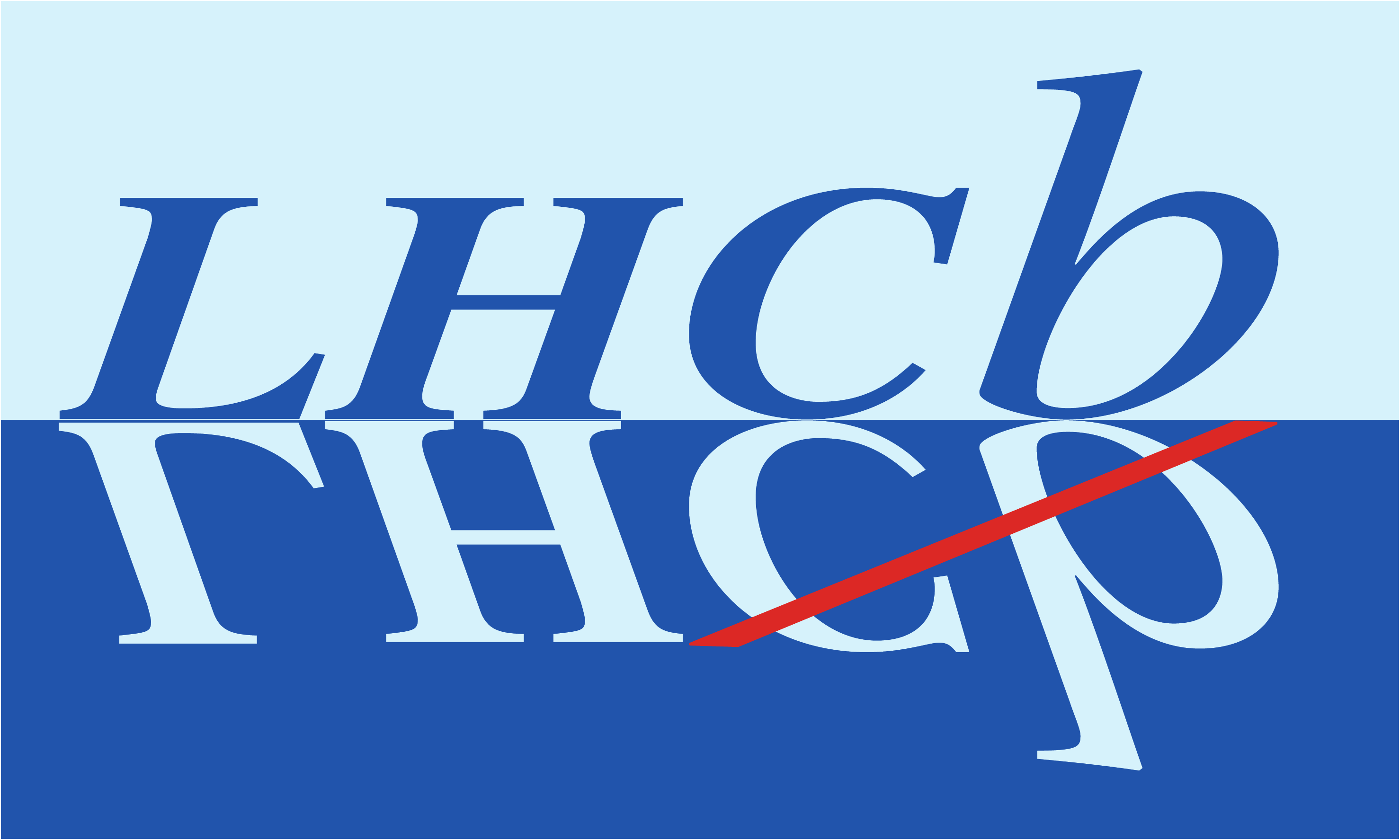}} & &}%
{\vspace*{-1.2cm}\mbox{\!\!\!\includegraphics[width=.12\textwidth]{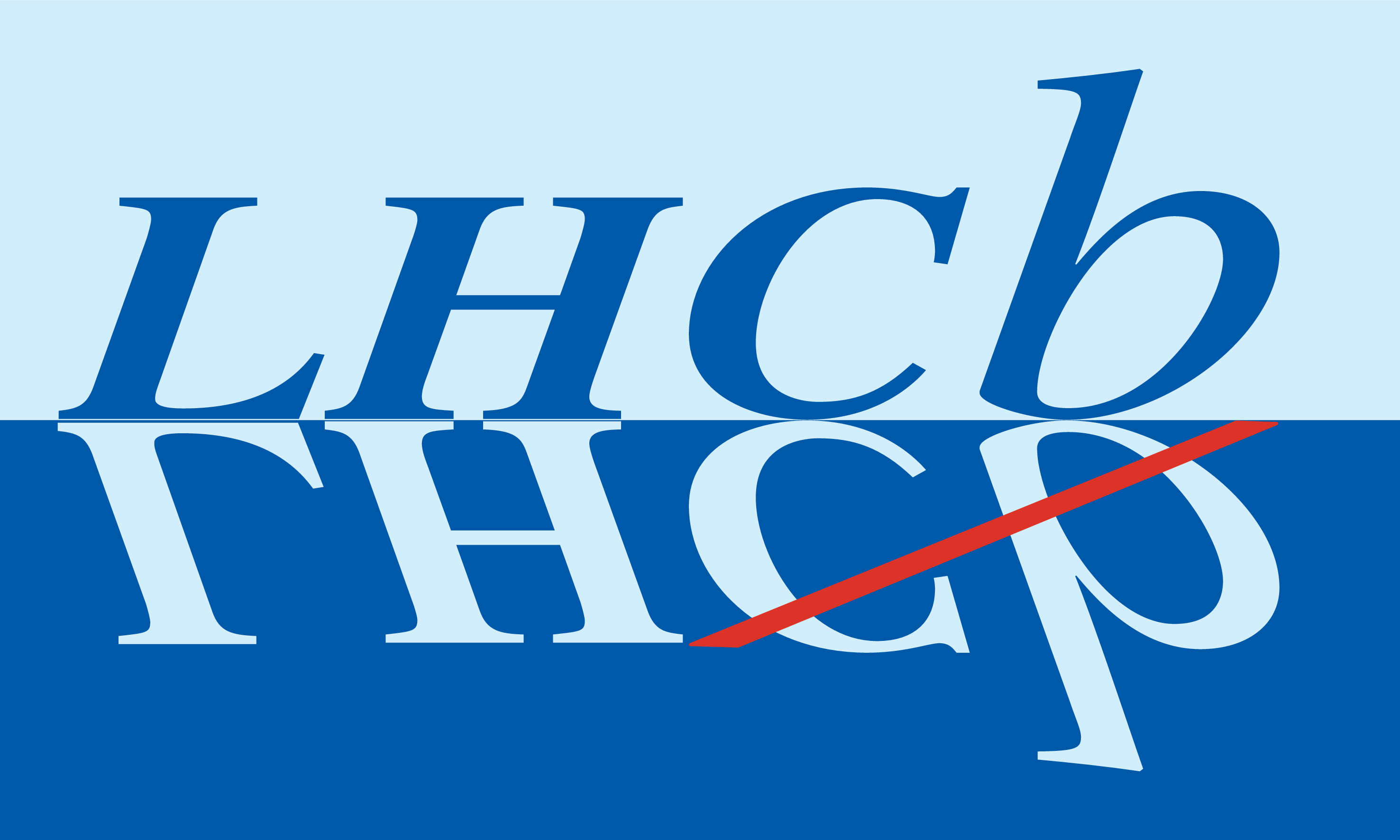}} & &}%
\\
 & & CERN-EP-2022-161 \\  
 & & LHCb-PAPER-2022-013 \\  
 & & \today \\ 
 & & \\
\end{tabular*}

\vspace*{2.0cm}

{\normalfont\bfseries\boldmath\huge
\begin{center}
  \papertitle 
\end{center}
}

\vspace*{1.0cm}

\begin{center}
\paperauthors\footnote{Authors are listed at the end of this Letter.}
\end{center}

\vspace{0.5cm}

\begin{abstract}
  \noindent

Jet fragmentation functions are measured for the first time in proton-proton collisions for charged pions, kaons, and protons within jets recoiling against a $Z$ boson. The charged-hadron distributions are studied longitudinally and transversely to the jet direction for jets with transverse momentum  20 $< \pt < 100$~GeV and in the pseudorapidity range $2.5 < \eta < 4$. The data sample was collected with the \lhcb experiment at a center-of-mass energy of 13 TeV, corresponding to an integrated luminosity of 1.64 fb$^{-1}$. Triple differential distributions as a function of the hadron longitudinal momentum fraction, hadron transverse momentum, and jet transverse momentum are also measured for the first time. This helps constrain transverse-momentum-dependent fragmentation functions. Differences in the shapes and magnitudes of the measured distributions for the different hadron species provide insights into the hadronization process for jets predominantly initiated by light quarks.

\end{abstract}

\vspace*{2.0cm}

\begin{center}
  Published in 
  Phys.~Rev.~D 108 (2023) L031103
\end{center}

\vspace{\fill}

{\footnotesize 
\centerline{\copyright~\papercopyright. \href{\paperlicenceurl}{\paperlicence}.}}
\vspace*{2mm}

\end{titlepage}

\newpage
\setcounter{page}{2}
\mbox{~}

\renewcommand{\thefootnote}{\arabic{footnote}}
\setcounter{footnote}{0}

\cleardoublepage

\pagestyle{plain} 
\setcounter{page}{1}
\pagenumbering{arabic}


Quarks and gluons can never be observed in isolation due to confinement in quantum chromodynamics (QCD). Thus, one of the challenges of QCD lies in relating the quark and gluon degrees of freedom of the theory to the bound-state hadrons observed in nature. A great deal of effort over the past several decades has gone into mapping out nucleon structure in terms of its quark and gluon constituents. A particular focus, in recent years, has been on the three-dimensional imaging of the nucleon~\cite{Anselmino:2020vlp,Kumericki:2016ehc}. Studying the mechanisms by which colored quarks and gluons hadronize into new color-neutral bound states offers complementary information connecting colored and hadronic degrees of freedom.  
 
In the standard collinear perturbative QCD factorization framework, single-inclusive hadron production in proton-proton ($pp$) collisions factorizes into the short-distance hard scattering of partons and the long-distance dynamics described by fragmentation functions (FFs) and parton distribution functions (PDFs). The latter parametrizes proton structure as a function of momentum fraction carried by a parton of an incoming proton taking part in the hard scattering process. Hadronization of charged particles is described by collinear FFs, denoted as $D_{q}^{h}(\z)$, where \z is the longitudinal momentum fraction of an outgoing parton $q$ carried by a produced hadron $h$ (see Ref.~\cite{Metz:2016swz} for a review of FFs). The FFs and PDFs are not fully calculable perturbatively and must be constrained by experimental measurements. In Monte Carlo (MC) generators, phenomenological models tuned to data are used to perform hadronization.~\cite{Artru:1974hr,Andersson:1983ia,Fox:1979ag}. Jet fragmentation functions (JFFs) are experimental observables describing jet substructure that measure the longitudinal momentum fraction carried by a hadron of a jet~\cite{Procura:2009vm,Kaufmann:2015hma,Chien:2015ctp,Neill:2016vbi,Kang:2016ehg,Kang:2017glf,Kang-tmdff-2019}. Within the soft-collinear effective theory framework, JFFs are constructed such that they can probe the standard collinear FFs, defined for inclusive single-hadron production with no requirement of a reconstructed jet. Similarly, transverse-momentum-dependent (TMD) JFFs defined within the soft-collinear effective theory framework can access standard TMD FFs~\cite{Collins:2011zzd}, traditionally measured in $e^+e^-$ collisions~\cite{Belle:2005dmx,Belle:2008fdv,BaBar:2013jdt,BaBar:2015mcn,BESIII:2015fyw,Belle:2019ywy} and semi-inclusive deep inelastic lepton-nucleon scattering~\cite{HERMES:2012uyd,COMPASS:2017mvk}. In addition to the dependence on the longitudinal momentum fraction \z, TMD FFs also depend on \jt, the transverse momentum of the produced hadron with respect to the jet axis in the case of a fully reconstructed jet, or the thrust axis in $e^+e^-$ collisions (see \eg Ref.~\cite{Belle:2019ywy}). Singly differential TMD JFFs for unidentified hadrons have previously been measured in proton-proton collisions at the LHC~\cite{ATLAS:2011myc,ATLAS:2019rqw,LHCb-PAPER-2019-012,ALICE:2020pga}. The excellent hadron identification capabilities at \lhcb allow for measurements of the JFFs for different particle species.

This Letter presents the first measurements of JFFs for identified charged hadrons in jets produced in association with a $Z$ boson in the forward region of $pp$ collisions. The main observables are the longitudinal momentum fraction of the jet carried by the hadron, \z, and the transverse component of the hadron momentum with respect to the jet axis, \jt, as found in Ref.~\cite{LHCb-PAPER-2019-012,STAR:2017akg} and defined as

\begin{equation}
\z = \frac{\phad \cdot \pjet}{|\pjet|^2}, \qquad \jt = \frac{| \phad \times \pjet |} {|\pjet|},
\label{eq:vars}
\end{equation}
where \phad and \pjet are the hadron and jet three-momentum vectors, respectively.

The dominant leading order hard process for $Z$+jet production in the LHCb acceptance is $q g \rightarrow Z q$ due to the asymmetry between the gluon and quark momentum fractions, verified with \pythia~8~\cite{Sjostrand:2007gs,*Sjostrand:2006za}, which enhances jets initiated by light valence quarks and provides sensitivity to the quark TMD FFs.

The JFFs measured using $Z$-tagged jets in this Letter are defined in terms of differential cross-sections $\deriv \sigma$ as 

\begin{equation}
f(\z,\jt) = \frac{\deriv \sigma}{\deriv \PScal \, \deriv \z \,\deriv \jt} \bigg/ \frac{\deriv \sigma}{\deriv \PScal},
\label{eq:jff1}
\end{equation}
\begin{equation}
F(\z) = \int \deriv \jt \;f(\z,\jt) = \frac{\deriv \sigma}{\deriv \PScal \, \deriv \z} \bigg/ \frac{\deriv \sigma}{\deriv \PScal},
\label{eq:jff2}
\end{equation}
\begin{equation}
F(\jt) = \int \deriv \z \;f(\z,\jt) = \frac{\deriv \sigma}{\deriv \PScal \, \deriv \jt} \bigg/ \frac{\deriv \sigma}{\deriv \PScal},
\label{eq:jff3}
\end{equation}
where the phase space $\deriv\PScal$ depends on the pseudorapidity of the $Z$ boson and the jet, and the vector sum and the difference between the transverse momenta of the $Z$ boson and the jet~\cite{Kang-tmdff-2019}.

The TMD JFF defined in Eq.~(\ref{eq:jff1}) is integrated over \jt to obtain the collinear JFF shown in Eq.~(\ref{eq:jff2}). The transverse profile is obtained by integrating the TMD JFF over \z as defined in Eq.~(\ref{eq:jff3}). Experimentally, these quantities can be expressed in terms of yields corrected for detector effects as
\begin{equation}
\hskip-0.3cm
f(\z, \jt) = \frac{1}{N_{\zjet}}\frac{\deriv \Nhad (\z, \jt)}{\deriv \z \, \deriv \jt},\; F(\z) = \frac{1}{N_{\zjet}}\frac{\deriv \Nhad (\z)}{\deriv \z},\; F(\jt) = \frac{1}{N_{\zjet}}\frac{\deriv \Nhad (\jt)}{\deriv \jt},
\label{eq:obsexp}
\end{equation} 
where \Nhad is the number of hadrons in $Z$-tagged jets for given \z and \jt, and $N_{\zjet}$ is the number of $Z$+jet pairs that contain charged hadrons.


The \lhcb detector~\cite{LHCb-DP-2008-001,LHCb-DP-2014-002} is a single-arm forward spectrometer covering the pseudorapidity range 2 $< \eta <$ 5. The detector includes a high-precision tracking system consisting of a silicon-strip vertex detector (VELO)~\cite{LHCb-DP-2014-001} surrounding the $pp$ interaction region, a silicon-strip detector located upstream of a dipole magnet with a bending power of about 4~Tm, and three stations of silicon-strip detectors and straw drift tubes~\cite{LHCb-DP-2013-003,LHCb-DP-2017-001} placed downstream of the magnet. The momentum resolution of charged particles provided by the tracking system is $\delta p/p \sim $ 0.5\% at low momentum and reaches 1.0\% at 200~GeV\footnote{In this article, natural units ($c$ = $\hbar$ = 1) are used.}. The VELO allows reconstruction of multiple primary vertices (PVs) and rejection of events with more than one PV or additional low-momentum tracks. 
Muons are identified by a system composed of alternating layers of iron and multiwire proportional chambers~\cite{LHCb-DP-2012-002}. Photons, electrons, and hadrons are distinguished by a calorimeter system consisting of scintillating-pad and preshower detectors, an electromagnetic calorimeter, and a hadronic calorimeter. Different types of charged hadrons are identified using information from two ring-imaging Cherenkov (RICH) detectors~\cite{LHCb-DP-2012-003}, with RICH 1 (C$_{4}$F$_{10}$ radiator) covering 2 to 60~GeV and RICH 2 (CF$_{4}$) covering 15 to 100~GeV. Simulated $pp$ collisions are generated using \pythia~8~\cite{Sjostrand:2007gs,*Sjostrand:2006za} with a specific \lhcb configuration~\cite{LHCb-PROC-2010-056}. Decays of hadronic particles are described by \evtgen~\cite{Lange:2001uf}, in which final-state radiation is generated using \photos~\cite{davidson2015photos}. Finally, the \geant toolkit~\cite{Agostinelli:2002hh,*Allison:2006ve} is used to simulate the interactions of the particles with the detector, as described in Ref.~\cite{LHCb-PROC-2011-006}. 

The data sample used in this analysis corresponds to an integrated luminosity of 1.64~fb$^{-1}$ collected at $\sqrt{s}$ = 13~TeV with the LHCb detector in 2016. The online event selection is performed by the muon trigger system, where $Z$ boson candidates are selected via their decay into two oppositely charged muons. The two muons are required to have $\pt > 20$~GeV, $2.0 < \eta(\mu) < 4.5$, and their invariant mass within the range $60 < M_{\mu\mu} < 120$~GeV, as applied in Ref.~\cite{LHCb-PAPER-2013-058,LHCb-PAPER-2019-012}. The muons must satisfy the track-reconstruction and muon-identification criteria applied in Ref.~\cite{LHCb-PAPER-2021-037}.
Jet reconstruction is performed offline using a particle-flow algorithm~\cite{LHCb-PAPER-2013-058}, where the neutral and charged candidates are clustered using the anti-$k_T$ algorithm~\cite{Cacciari:2008gp,Cacciari:2011ma} with the $R$ parameter set to 0.5. The selection criteria for $Z$+jet pairs and tracks inside the jets closely follow those described in Ref.~\cite{LHCb-PAPER-2019-012}. The fiducial criteria require that the jet with the highest \pt in the event, which is analyzed for these measurements, has $20 < \pt(\textrm{jet}) < 100$~GeV and $2.5 < \eta(\textrm{jet}) < 4.0$. Additional jets with $\pt(\textrm{jet}) > 15$~GeV are used in unfolding detector effects. The tighter $\eta(\textrm{jet})$ requirement ensures that all the jet constituents are contained within the detector acceptance. To reduce the rate of jets associated with a different primary vertex than the $Z$ candidate, only events with a single reconstructed primary vertex are analyzed. The jets must be well separated from the $Z$ candidate by requiring an azimuthal separation $\Delta \phi_{\Z-\textrm{jet}}$ greater than $\frac{7\pi}{8}$, and are rejected if one of the muons is found within $\Delta R = \sqrt{\Delta \eta^2 + \Delta \phi^2} < 0.5$, defined with respect to the jet momentum. The charged hadron candidates must be constituents of the jet, fall within $\Delta R < 0.5$ of the jet, and have a good quality track with a minimum \pt (\p) of 0.25 (4) GeV.

The number of $Z$+jet pairs in each jet \pt interval, used in the normalization of the JFFs, is corrected to account for reconstruction and selection efficiencies. The same correction factors are applied to the hadron distributions in jets. The muon detection and reconstruction efficiencies are determined in data using the tag-and-probe method employed in the inclusive $Z$ boson cross-section measurements of LHCb~\cite{LHCb-PAPER-2015-049,LHCb:2015okr,LHCb-PAPER-2021-037}. The efficiency to reconstruct and identify the jet in the event is evaluated from simulation. This efficiency increases with \pt, from $\approx 85\%$ for jets with \pt of 20~GeV to saturate at $\approx 95\%$ for jets with \pt of 30~GeV and above.

The charged hadron candidates inside the reconstructed jets are identified by the particle-identification systems~\cite{LHCb-DP-2008-001,LHCb-DP-2012-003}. Reconstructed charged hadron yields are corrected for the track-reconstruction efficiency, effects from misreconstructed tracks or false association with jets, and particle misidentification on a track-by-track basis. Simulation is used to determine the track reconstruction efficiency for pions, kaons, and protons separately as a function of momentum and pseudorapidity. The probability of hadronic interactions in the detector material of $\approx$20\% $\lambda$ (nuclear interaction lengths), and decays in flight, results in track reconstruction efficiencies of 79\%, 77\% and 63\% for pions, kaons, and protons, respectively. 

The particle identification (PID) efficiency
is determined in intervals of particle momentum, pseudorapidity, and track multiplicity using dedicated data control samples~\cite{LHCb-PUB-2016-021}. The (mis)identification probabilities of charged hadrons are derived from these samples and used to construct a PID matrix. The particle misidentification effects are unfolded by solving  
\begin{equation}
\mathbf{x}^{ \,\mathrm{rec}} = A \; \mathbf{x}^{ \,\mathrm{unf}}~ 
\label{eq:pidunfold}
\end{equation}
in each momentum interval. The vector $\mathbf{x}^{\,\mathrm{unf}}$ represents the unfolded yields of the three particles species $\pipm,\Kpm$, and $p^{\pm}$ in a given momentum interval and $\mathbf{x}^{\,\mathrm{rec}}$ the corresponding reconstructed yields at detector level. A matrix element $A_{\; ij}$ represents the probability of a particle $j$ to be reconstructed as $i$. The probabilities are weighted to describe the pseudorapidity and track-multiplicity distributions in data.  Correction factors for pions (kaons) ranged from 0.98 (0.7) to 1.1 (1.03) as a function of momentum. For protons, the corrections were as large a factor of 4 at momentum below 10 GeV where PID is not fully efficient and ranged from 0.5 to 1.2 above 10 GeV. The PID corrections for kaons and protons dominate at the lowest momentum, while they compete with the track reconstruction efficiencies at higher momentum. 

The uncertainties on the resulting PID-unfolded momentum distributions $\mathbf{x}^{\,\mathrm{unf}}$ are estimated using a bootstrap method~\cite{Efron:bootstrap} with 500 trials and are statistically dominated. The impurity of the charged-hadron sample due to misidentification of nonhadronic particles or long-lived hyperons as charged pions, kaons, or protons is less than 5\% of the statistical uncertainties in all momentum intervals. The particle-species-dependent efficiencies can be biased if a particle is misidentified. The efficiencies are corrected for these effects after the PID unfolding.

The unfolding of detector effects in the distributions of transverse momentum and pesudorapidity of jets in $Z$+jet pairs is performed using machine learning (ML) techniques based on the iterative unbinned Bayesian unfolding method as implemented in Ref.~\cite{Andreassen:2019cjw}. In this method, deep neural networks are employed as classifiers to estimate likelihood ratios that are used to update event weights in simulation. The unfolding method is validated with a closure test using the simulated data. In this analysis, two iterations are performed based on the best results of the closure tests. The unfolded jet-\pt distributions are consistent within statistical uncertainties between the ML method employed in this analysis and the binned iterative Bayesian method~\cite{IBU,Adye:2011gm} with two iterations. The charged hadron distributions inside jets in $\z, \jt$, and their joint distributions, are unfolded simultaneously with the jet transverse momentum and pseudorapidity. 


Sources of systematic uncertainties on the jet transverse-momentum distribution of $Z$+jet pairs and identified charged hadrons in $\z$ and $\jt$ are evaluated. The former arise from the background contributions from fake jets (0.2\%) and incorrectly reconstructed $Z$ bosons (1.4\%) due to hadrons misidentified as muons. Additionally, uncertainties on the jet reconstruction are determined by comparing jet-quality quantifiers between the data and simulation (1.8\%), similar to the method employed in Ref.~\cite{LHCb-PAPER-2016-011}. Uncertainties related to the muon reconstruction efficiencies are found to be negligible. The $Z$+jet selection and jet reconstruction uncertainties added in quadrature return a total uncertainty of 2.4\%. 

The primary uncertainties associated with the simulated detector response of the jets arise from the jet-energy scale and resolution. The jet-energy scale has been studied in previous measurements of the $Z$+jet cross-section~\cite{LHCb-PAPER-2016-011,LHCb-PAPER-2013-058}. Exploiting the \pt balance between the $Z$ boson and a single recoiled jet in the event, the uncertainty on the jet-energy scale is determined to be 3\%. The uncertainties on the fragmentation measurements are estimated by repeating the analysis with the energy scale in the simulation varied by one standard deviation and taking the difference in the distributions, as in Refs.~\cite{LHCb-PAPER-2019-012,LHCb-PAPER-2016-011,LHCb-PAPER-2013-058}. Similarly, the systematic uncertainty due to the jet-energy resolution is evaluated by independently varying each component of the reconstructed jet momentum in the simulation by the uncertainty on the jet resolution. This procedure is repeated until the difference in the unfolded distributions between the nominal and smeared jet momentum stabilizes.

The sources of systematic uncertainties on the identified charged-hadron distributions include the tracking efficiency and particle identification. The effects of the statistical precision of the efficiency are evaluated by smoothing the two-dimensional efficiency and repeating the analysis. The differences in the distributions between the smoothed and nominal efficiencies are taken as uncertainties on the tracking efficiency. Additionally, PID-dependent uncertainties attributed to the uncertainties on the material budget implemented in the simulation are found to be 1.50\%, 1.27\% and 3.3\% for pions, kaons, and protons, respectively. The uncertainty on the track-selection requirement to remove spurious tracks formed by accidentally matched detector hits and charged hadrons not associated with the jet is negligible. For identified charged-hadron distributions, the systematic uncertainties on the PID are determined from the uncertainties on the PID-unfolded momentum distributions.

The systematic uncertainties on the unfolding method are determined by taking the standard deviation of weighted absolute differences in the ratio between the unfolded and generated distributions. The uncertainties on the unfolded number of jets and the non-normalized hadronization variables are 1.1\% and 0.8\%, respectively.

\begin{figure}[tbh!]
  \begin{center}
\includegraphics[width=1.0\linewidth]{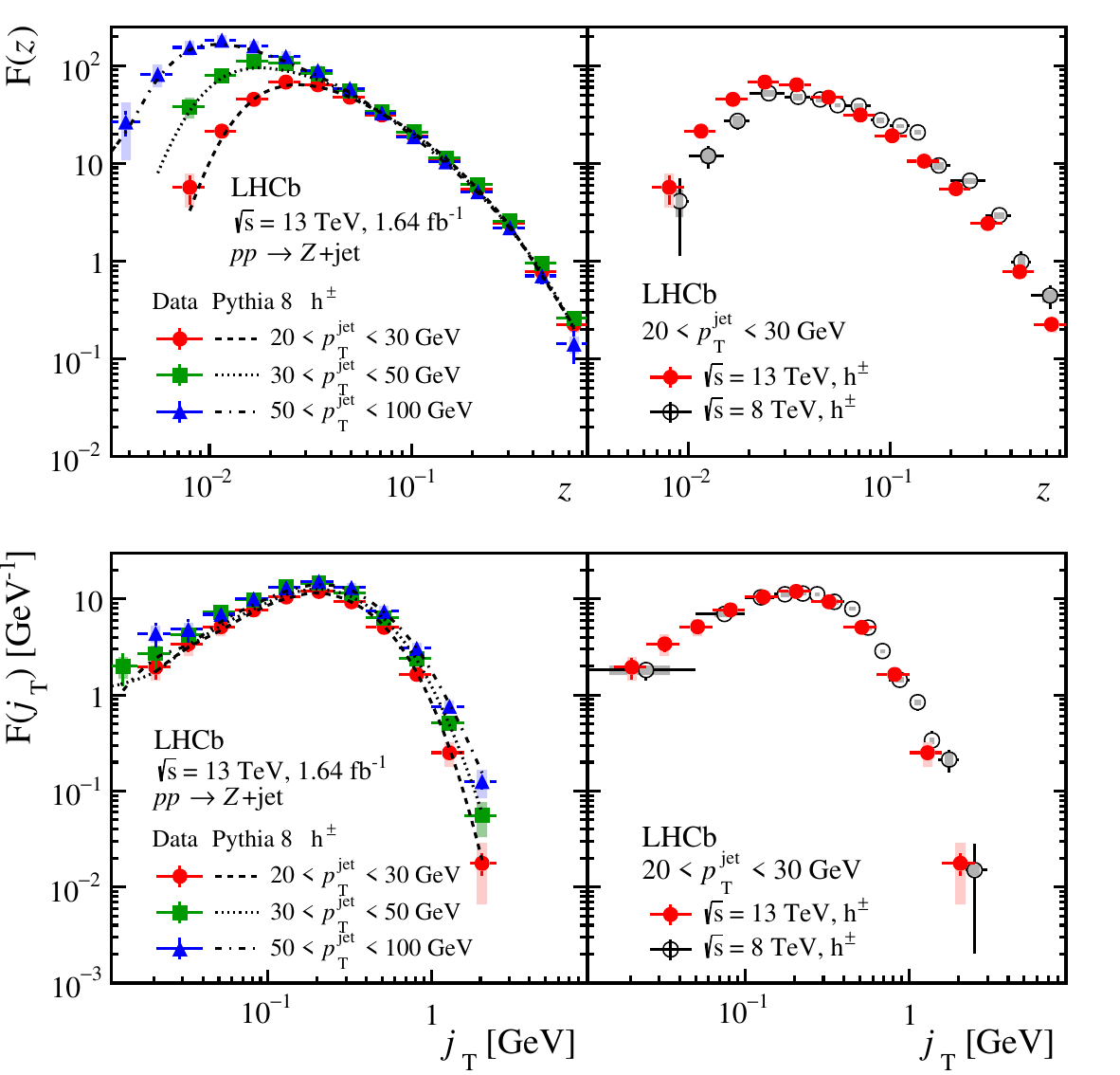}
    \vspace*{-0.5cm}
  \end{center}
  \caption{
    \small {Distributions of (top) the longitudinal momentum fraction and (bottom) the transverse momentum of charged hadrons (pions, kaons, and protons combined) with respect to the jet axis in three jet \pt intervals and (right) comparisons with previous results at $\sqrt{s} = 8$~TeV for jets with 20 $<\pt<$ 30~GeV~\cite{LHCb-PAPER-2019-012}. Statistical (systematic) uncertainties are shown in bars (boxes). }
    }
  \label{fig:nonid}
\end{figure}

Figure~\ref{fig:nonid} shows \z and \jt distributions in three intervals of jet \pt for unidentified charged hadrons. The \z distributions show a humpbacked structure in \z $< 0.04$ due to both color coherence and kinematic requirements. Color coherence is a manifestation of parton hadron duality~\cite{Azimov:1985by} where the perturbative partonic structure of a jet defines the momenta of the final hadrons, and produces this structure due to the suppression of wide-angle gluon emissions. In these distributions the kinematic requirements on the tracks shift the peak of the structure to varying degrees for different jet \pt intervals; higher \pt jets can probe smaller \z. At mid-to-high $\z$, $0.04 < \z < 0.4$, scaling behavior is seen across all jet \pt intervals. An overall increase of particle production in all regions of \jt for jets with higher \pt is seen, with a larger increase at high \jt. Comparisons with previous measurements at $\sqrt{s} = 8$~TeV~\cite{LHCb-PAPER-2019-012} show a general similarity in shape. While the measurements indicate that charged hadron production may be slightly shifted towards lower $\z$ at $\sqrt{s} = 13$~TeV for a given jet \pt, the \jt distributions are consistent within uncertainties between the two center-of-mass energies.  

\begin{figure}[bth!]
  \begin{center}
\includegraphics[width=1.\linewidth]{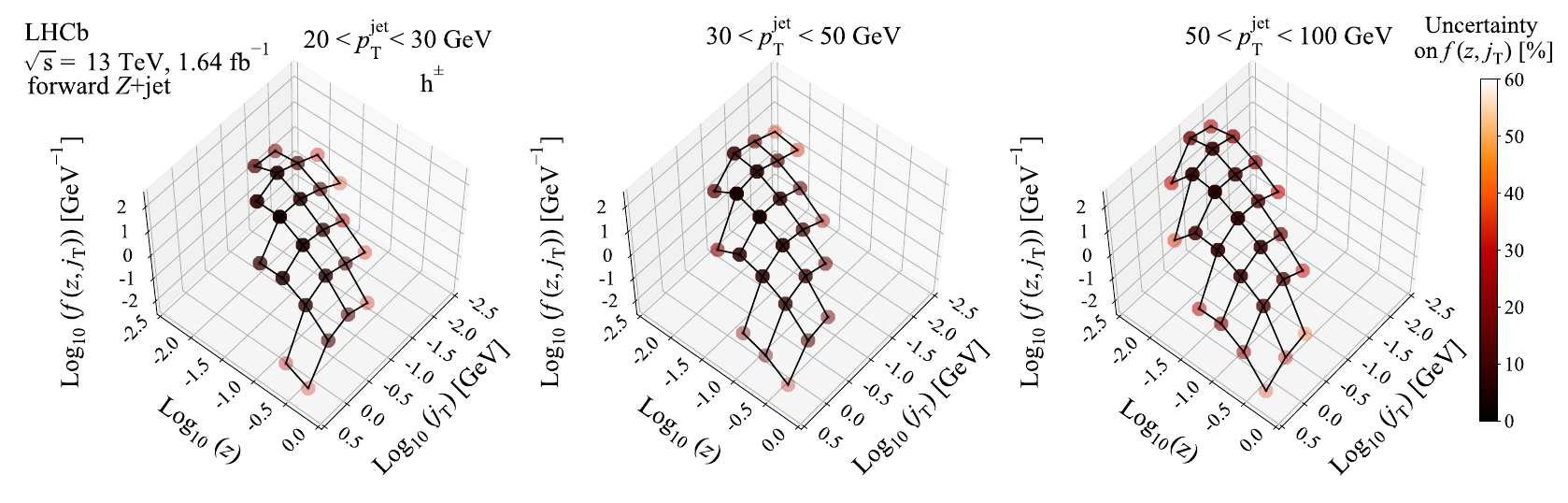}
    \vspace*{-0.5cm}
  \end{center}
  \caption{ \small{ Double differential JFFs of the longitudinal momentum fraction \z and the transverse momentum \jt of charged hadrons (pions, kaons, and protons combined) in three jet \pt intervals. 
  }
    }
  \label{fig:hadron3d}
\end{figure}

The double differential JFFs in \jt, \z, in three jet \pt intervals for unidentified charged hadrons are shown in Fig.~\ref{fig:hadron3d}. Charged hadrons carrying a larger momentum fraction along the jet axis tend to have a larger transverse momentum with respect to the jet axis. With increasing jet \pt, the centroid of the joint distributions moves towards a smaller \z, a region dominated by soft particle production, and a larger $\jt$, resulting in wider jets. Charged particles also carry a larger \jt for a given \z in jets with a higher \pt. This is consistent with Markov chain fragmentation models, \eg the string~\cite{Andersson:1983ia} or cluster model~\cite{Fox:1979ag}, where a momentum kick transverse to the parton system is sampled independently per hadron. Jets with higher \pt experience longer Markov chains, resulting in a higher \jt for a given \z.

\begin{figure}[htb]
  \begin{center}
\includegraphics[width=1.\linewidth]{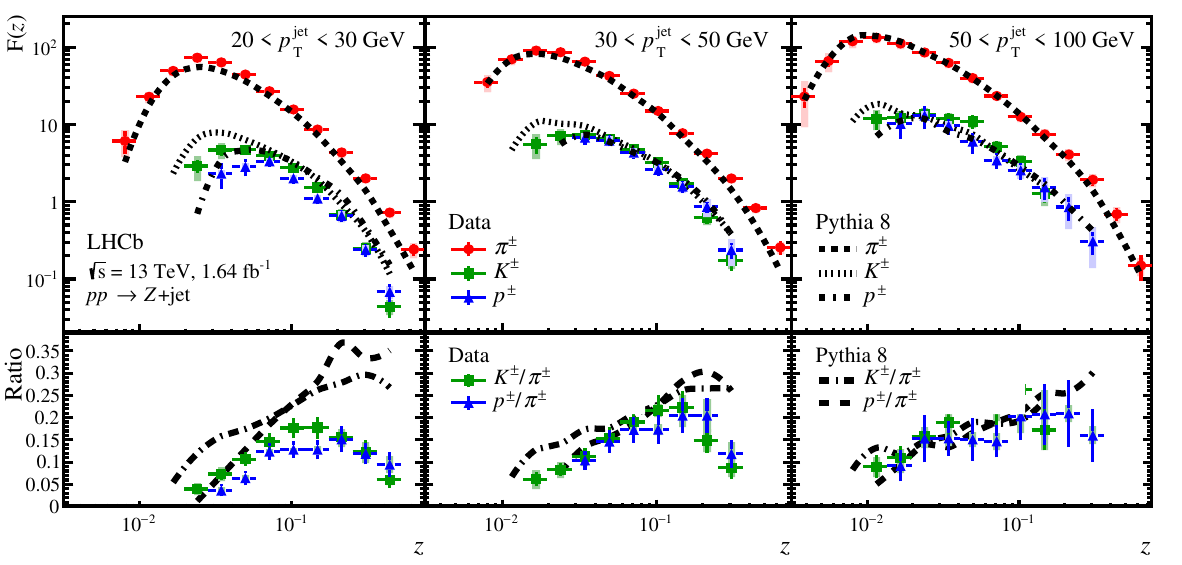}
    \vspace*{-0.5cm}
  \end{center}
  \caption{
\small{Collinear jet fragmentation functions of (top) identified pions, kaons and protons in three jet \pt intervals and (bottom) the ratios of kaons to pions and protons to pions. Statistical (systematic) uncertainties are shown in bars (boxes). }
    }
  \label{fig:pidratio}
\end{figure}

The \z distributions for identified charged hadrons and the ratios of heavier particles with respect to pions are shown in Fig.~\ref{fig:pidratio}. Pions are the predominant charged hadron produced due to their low mass and the flavor content of the initial-state protons. Hadrons with heavier mass require a larger \z threshold for their formation, leading to the position of the maximum at a higher \z. In the lowest jet \pt interval, proton production relative to kaon production is clearly suppressed for lower \z values. When different jet \pt intervals are overlaid, the scaling behavior across all jet \pt intervals begins at $\z\sim 0.07$ for heavier particles and 0.03 for pions.

The JFFs and the ratios are compared to predictions from \pythia8 in Fig.~\ref{fig:nonid} and Fig.~\ref{fig:pidratio}. The predictions are generated using \pythia~8.186 with the CT09MCS PDF set and a specific \lhcb configuration~\cite{LHCb-PROC-2010-056}. In general, \pythia~8 describes unidentified charged hadron distributions well with only slight underestimation while the number of charged pions (kaons and protons) are largely underestimated (overestimated). The production of heavier particles relative to pions is well described by \pythia~8 at high jet \pt, while at low  jet \pt \pythia~8 significantly overestimates it. These data can be used to tune MC generators for production of identified charged particles. 

\begin{figure}[bth!]
  \begin{center}
\includegraphics[width=1.\linewidth]{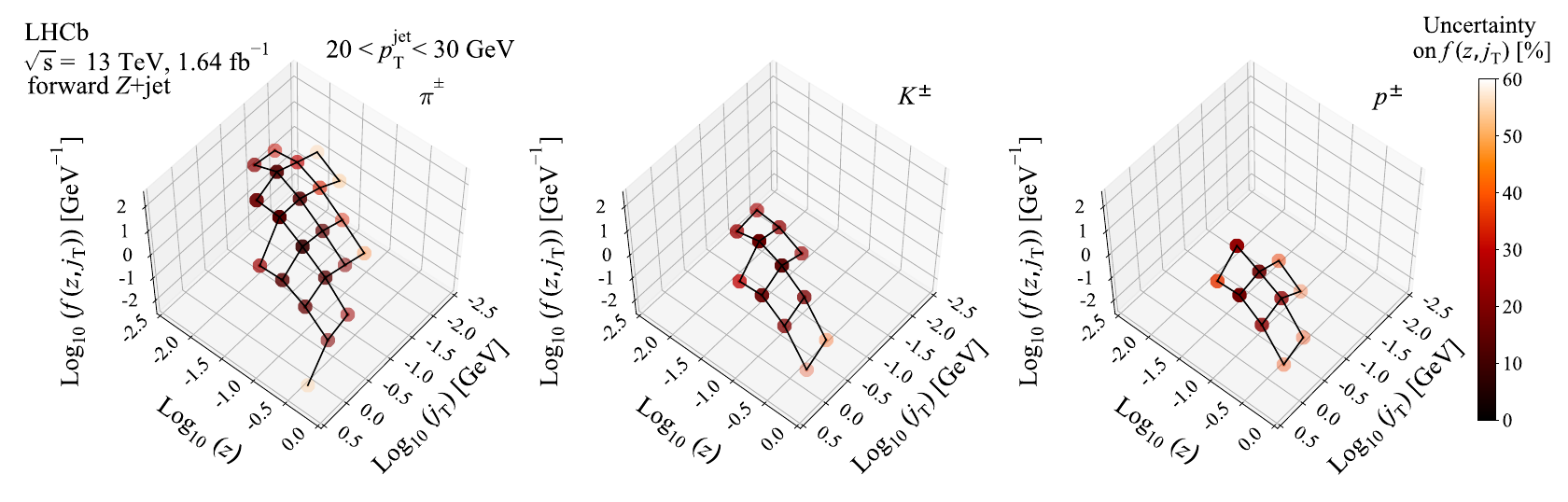}
    \vspace*{-0.5cm}
  \end{center}
  \caption{\small{Joint distributions of the longitudinal momentum fraction \z and the transverse momentum \jt of identified charged (left) pions, (middle) kaons and (right) protons in jets with 20 $< \pt <$ 30~GeV. }
    }
  \label{fig:pid3d}
\end{figure}

Figure~\ref{fig:pid3d} shows the TMD JFFs measured as joint distributions in \z and \jt for the three separate particle species. The center of the distribution shifting towards higher values in both \z and \jt with the mass of the particle suggests that heavier hadrons are produced from harder partons.   

In summary, the LHCb collaboration has measured the joint distributions in two kinematic variables simultaneously, probing the longitudinal and transverse profiles of identified charged pions, kaons, and protons inside predominantly light-quark-initiated jets for the first time. These distributions describe the 3D picture in the collinear and transverse dimension with respect to the jet axis, and the hadron-mass hierarchy in the hadronization processes. They will help constrain TMD FFs in uncharted phase space. These measurements exploit the full particle-identification capabilities of the \lhcb detector. The joint distributions for all charged hadrons have also been measured for the first time.

The collinear JFFs for identified charged hadrons exhibit the effects of quark-flavor content inside the proton. The relative jet-fragmentation functions of heavier particles to pions could provide insights into the role of the valence versus sea quarks in the parton shower leading to hadronization. The projected single-variable distributions in $\jt$ and $\z$ have been measured for identified and all charged hadrons and compared to the previous \lhcb results at 8~TeV. Overall, similar patterns are seen in $\jt$ and $\z>0.04$ between $\sqrt{s} =$ 8~TeV and 13~TeV.

\section*{Acknowledgements}

\noindent We express our gratitude to our colleagues in the CERN
accelerator departments for the excellent performance of the LHC. We
thank the technical and administrative staff at the LHCb
institutes.
We acknowledge support from CERN and from the national agencies:
CAPES, CNPq, FAPERJ and FINEP (Brazil); 
MOST and NSFC (China); 
CNRS/IN2P3 (France); 
BMBF, DFG and MPG (Germany); 
INFN (Italy); 
NWO (Netherlands); 
MNiSW and NCN (Poland); 
MEN/IFA (Romania); 
MICINN (Spain); 
SNSF and SER (Switzerland); 
NASU (Ukraine); 
STFC (United Kingdom); 
DOE NP and NSF (USA).
We acknowledge the computing resources that are provided by CERN, IN2P3
(France), KIT and DESY (Germany), INFN (Italy), SURF (Netherlands),
PIC (Spain), GridPP (United Kingdom), 
CSCS (Switzerland), IFIN-HH (Romania), CBPF (Brazil),
Polish WLCG  (Poland) and NERSC (USA).
We are indebted to the communities behind the multiple open-source
software packages on which we depend.
Individual groups or members have received support from
ARC and ARDC (Australia);
Minciencias (Colombia);
AvH Foundation (Germany);
EPLANET, Marie Sk\l{}odowska-Curie Actions and ERC (European Union);
A*MIDEX, ANR, IPhU and Labex P2IO, and R\'{e}gion Auvergne-Rh\^{o}ne-Alpes (France);
Key Research Program of Frontier Sciences of CAS, CAS PIFI, CAS CCEPP, 
Fundamental Research Funds for the Central Universities, 
and Sci. \& Tech. Program of Guangzhou (China);
GVA, XuntaGal, GENCAT and Prog. Atracci\'on Talento, CM (Spain);
SRC (Sweden);
the Leverhulme Trust, the Royal Society
 and UKRI (United Kingdom).

\addcontentsline{toc}{section}{References}
\ifx\mcitethebibliography\mciteundefinedmacro
\PackageError{LHCb.bst}{mciteplus.sty has not been loaded}
{This bibstyle requires the use of the mciteplus package.}\fi
\providecommand{\href}[2]{#2}

\clearpage
\centerline
{\large\bf LHCb collaboration}
\begin
{flushleft}
\small
R.~Aaij$^{32}$\lhcborcid{0000-0003-0533-1952},
A.S.W.~Abdelmotteleb$^{50}$\lhcborcid{0000-0001-7905-0542},
C.~Abellan~Beteta$^{44}$,
F.~Abudin{\'e}n$^{50}$\lhcborcid{0000-0002-6737-3528},
T.~Ackernley$^{54}$\lhcborcid{0000-0002-5951-3498},
B.~Adeva$^{40}$\lhcborcid{0000-0001-9756-3712},
M.~Adinolfi$^{48}$\lhcborcid{0000-0002-1326-1264},
H.~Afsharnia$^{9}$,
C.~Agapopoulou$^{13}$\lhcborcid{0000-0002-2368-0147},
C.A.~Aidala$^{76}$\lhcborcid{0000-0001-9540-4988}
% ...

{\footnotesize \it

% ...
$^{9}$Universit{\'e} Clermont Auvergne, CNRS/IN2P3, LPC, Clermont-Ferrand, France\\
% ...
$^{13}$LPNHE, Sorbonne Universit{\'e}, Paris Diderot Sorbonne Paris Cit{\'e}, CNRS/IN2P3, Paris, France\\
% ...
$^{32}$Nikhef National Institute for Subatomic Physics, Amsterdam, Netherlands\\
% ...
$^{40}$Instituto Galego de F{\'\i}sica de Altas Enerx{\'\i}as (IGFAE), Universidade de Santiago de Compostela, Santiago de Compostela, Spain\\
% ...
$^{44}$Physik-Institut, Universit{\"a}t Z{\"u}rich, Z{\"u}rich, Switzerland\\
% ...
$^{48}$H.H. Wills Physics Laboratory, University of Bristol, Bristol, United Kingdom\\
% ...
$^{50}$Department of Physics, University of Warwick, Coventry, United Kingdom\\
% ...
$^{54}$Oliver Lodge Laboratory, University of Liverpool, Liverpool, United Kingdom\\
% ...
$^{62}$Syracuse University, Syracuse, NY, United States\\
% ...
$^{76}$University of Michigan, Ann Arbor, MI, United States, associated to $^{62}$\\
\bigskip
% ...
}
\end{flushleft}
\end{document}